# A DC magnetic metamaterial

F. Magnus[1], B. Wood[1], J. Moore[1], K. Morrison[1], G. Perkins[1], J. Fyson[2], M. C. K. Wiltshire[1,3], D. Caplin[1], L. F. Cohen[1], J. B. Pendry[1]

[1]Physics Department, Imperial College London, Exhibition Road, London SW7 2AZ, UK
[2]European Research, Kodak Limited, 332 Science Park, Milton Road, Cambridge CB4 0WN, UK
[3]Imaging Sciences Department, Imperial College London, Ducane Road, London W12 0NN, UK

**Electromagnetic metamaterials[1] are a class of materials which have been artificially structured on a subwavelength scale. They are currently the focus of a great deal of interest because they allow access to previously unrealisable properties like a negative refractive index[2]. Most metamaterial designs have so far been based on resonant elements, like split rings[3], and research has concentrated on microwave frequencies and above. In this work, we present the first experimental realisation of a non-resonant metamaterial designed to operate at zero frequency. Our samples are based on a recently-proposed template[4] for an anisotropic magnetic metamaterial consisting of an array of superconducting plates. Magnetometry experiments show a strong, adjustable diamagnetic response when a field is applied perpendicular to the plates. We have calculated the corresponding effective permeability, which agrees well with theoretical predictions. Applications for this metamaterial may include non-intrusive screening of weak DC magnetic fields.**

The first metamaterials[3,5] were designed to operate at microwave frequencies. Since then, while there has been some research on radio-frequency metamaterials[6], most of the research effort has been focused on higher frequencies: technologically-important microwaves or visible light[7]. The low-frequency end of the spectrum has remained relatively unexplored.

In addition, the majority of metamaterials devised to date consist of an arrangement of resonant components. There is a good reason for this: the response of a resonator varies greatly as a function of the frequency at which it is being driven. Close to the resonant frequency, the amplitude of the response can be very large, while the phase changes. The range of available values of the response function, or susceptibility, is therefore very wide. One of the crowning achievements of (and driving forces behind) metamaterials research is the realization of a negative refractive index[2,8], and a simple argument shows that this cannot be achieved without relying on resonant structures. However, the price of working close to the resonant frequency is that losses and frequency dispersion are greatest here. When a negative response is not required then a non-resonant structure is advantageous.

Another recent development means that there is new demand for metamaterials with non-negative anisotropic parameters. Transformation optics[9] is a design paradigm that allows a new level of control over electromagnetic fields. For a given design, it provides a recipe for the material parameters as a function of position. The parameters generated in this way are always non-negative and anisotropic. A spectacular demonstration of the technique was provided by the construction of an electromagnetic cloak[10] using metamaterials. The interior of the cloak is shielded from microwaves with minimal disruption to the exterior fields.

These ideas have been combined in a design for a non-resonant metamaterial with an anisotropic response to DC magnetic fields[4]. The metamaterial is composed of superconducting plates arranged in a lattice (Fig. 1). The superconducting elements exclude static magnetic fields, and provide the foundation for the diamagnetic effect. An applied field can still penetrate the structure, but is concentrated in the gaps between the plates. This has a negligible effect when a field is applied in a direction parallel to the plates, as long as the plates are sufficiently thin. The in-plane effective permeability is therefore close to unity. However, when a field is applied normal to the plates, the distortion of the field lines around the plates means that the metamaterial as a whole shows diamagnetic properties. The strength of the response when the field is perpendicular to the plates depends on the ratio between the dimension of the plates and the lattice spacing. The main point is that the result is a metamaterial with a highly tuneable, diamagnetic, anisotropic effective permeability. Our aim was to construct and measure the effective permeability of samples based on this template.

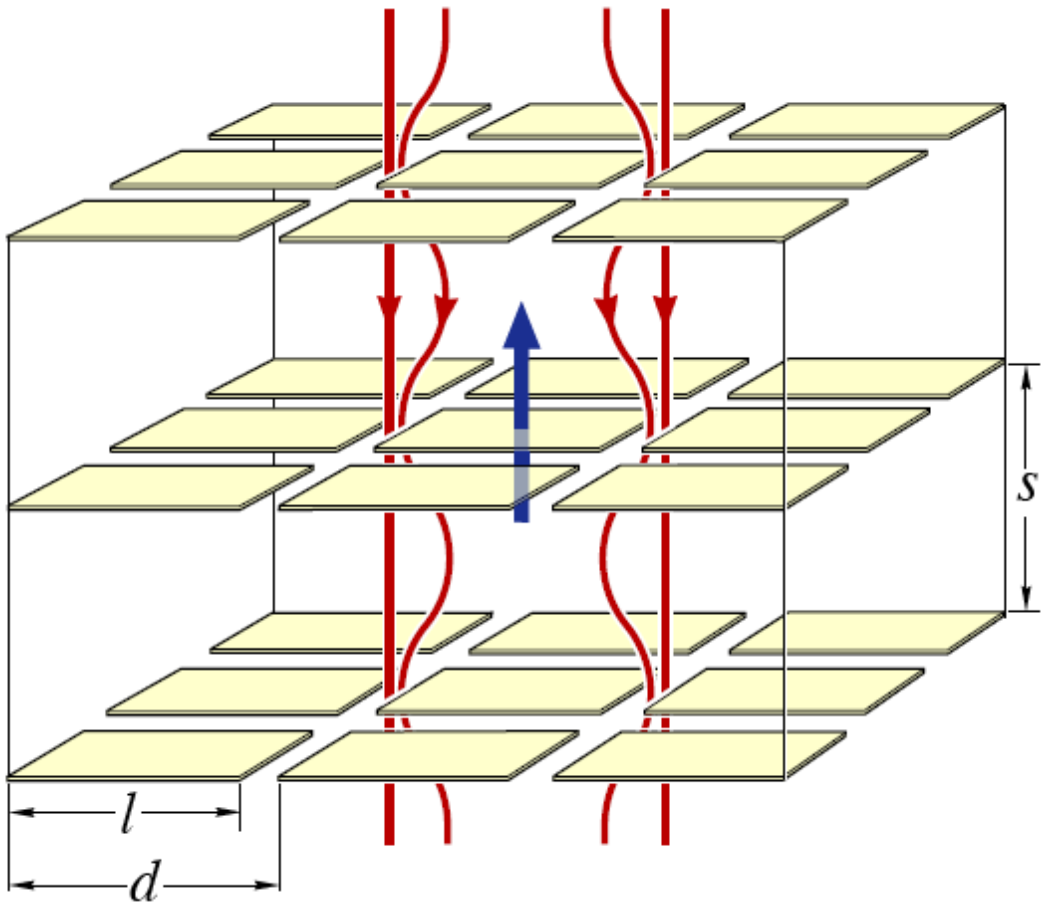

**Figure 1** Schematic showing the layers of square superconducting plates combined into a tetragonal lattice. The magnetic field lines are pushed into the gaps between the plates. The arrow through the central plate shows the induced magnetic dipole moment.

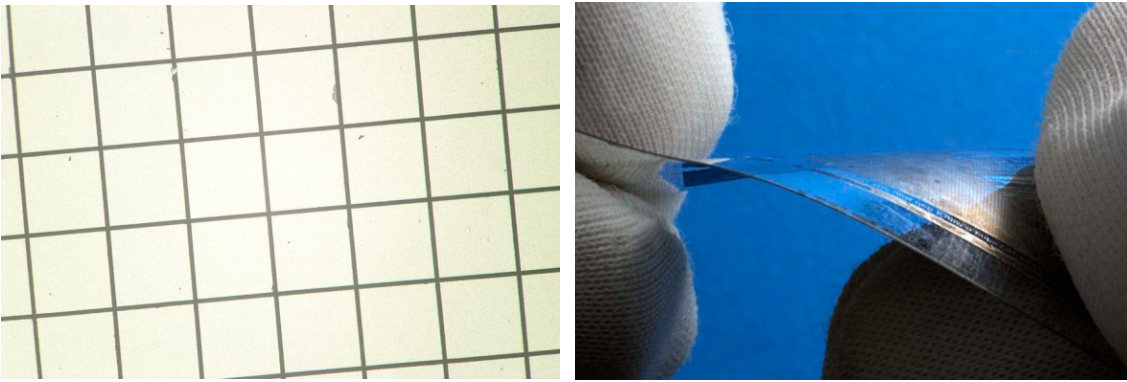

**Figure 2** Implementation of the metamaterial design. **a** An optical image of a single glass slide showing an array of 163 μm lead squares separated by 4 μm gaps. **b** The alternative PET substrate allows the samples to be flexible.

The natural way to assemble the metamaterial is by stacking layers of plates. In our original theoretical treatment, the calculations were performed for a set of plates that were precisely aligned vertically. For ease of manufacture we have extended the theoretical calculations to investigate the influence of misalignment of the layers. The study was performed by simulating a system with consecutive layers anti-aligned. Following our earlier work[4], we calculated the band structure (from which the effective medium parameters can be deduced) for the anti-aligned system, and found it to be unaffected by the misalignment for the geometry where the plate thickness is much less than the vertical plate separation

To qualify as a metamaterial, the lattice spacing must be significantly smaller than the wavelength of light. This is not a problem at zero frequency, where the wavelength diverges. The only size constraint was provided by the magnetometer, which would not accept samples with dimensions larger than 4 mm. The system under study must also contain some minimum number of cells of the order of ten: a single superconducting plate cannot sensibly be described as a metamaterial. This meant that the cells had to be small, with lattice spacing on the order of 100 μm. The thickness of the plates is also important; it should be small compared to the lattice spacing but large compared to the field penetration length of the superconductor, which is around 50 nm for our chosen superconductor, lead[11].

In accordance with these constraints, lead films of 300 nm thickness were thermally evaporated onto 100 μm thick glass slides and patterned into arrays of plates using a conventional photolithographic lift-off process. The plates were arranged in a 10×10 square lattice with a lattice constant of 167 μm and a spacing of 2–34 ± 0.2 μm. (A control pattern consisting of a solid 1.67×1.67 $mm^2$ square was also made.) After patterning, the slides were glued together to form a three-dimensional tetragonal lattice, with the patterns on different layers aligned to within 10 μm using an optical microscope. Finally, the samples were diced with a diamond saw to separate each sample. Figure 2a shows part of a single slide.

With future applications in mind, we also manufactured samples using a flexible PET substrate (Fig. 2b): silver was first evaporated onto the substrate through a mask, giving the same array of square tiles; electroplating was then used to add lead to the tiles to a thickness of a few microns.

The magnetic moment of each sample was measured using an Oxford Instruments split coil vibrating sample magnetometer (VSM) at 4.2 K as a function of applied magnetic field. The sample could be rotated about an axis perpendicular to the field to obtain the angular dependence of the magnetic moment. We performed minor magnetic moment versus field (*m–H*) loops to determine the lowest field at which irreversibility sets in. The full *m–H* loops reveal that the lead behaves like a weak type II superconductor with flux pinning, as would be expected for a disordered evaporated film. To ensure that flux pinning was not playing a role in the low field data we warmed the sample above the critical temperature after the minor loop routine, and then performed a virgin *m–H* curve being careful to remain in the reversible regime. Because the field between the plates is greatly enhanced (by a factor of order ), the applied field was therefore limited to around 10 Oe for the smallest gap size (4 $\mu$m). The magnet that we used allows field increments of 0.1 Oe, so the reversible Meissner slope was well resolved even in this instance. This slope, when normalized to the sample volume, yields the diamagnetic susceptibility $\chi$ and hence the effective permeability $\mu = 1+\chi$.

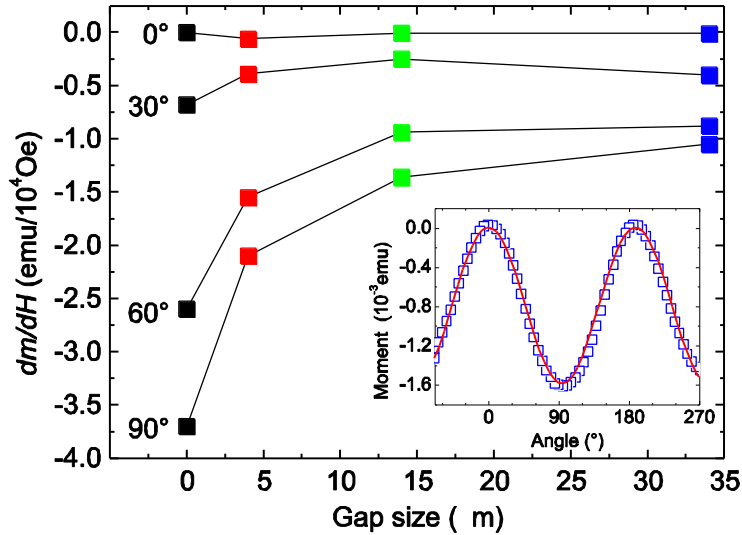

**Figure 3** The effective DC susceptibility d*m*/d*H* as a function of gap size and sample orientation. The results for the third sample (with gap size 14 $\mu$m) have been scaled to allow for the fact that this sample consists of only 6 layers while the others have 9. An angle of 90° means that the applied field is perpendicular to the layers within the sample. The inset more clearly shows the angular dependence of the response; the continuous line is a fit to the data, and is proportional to $sin^2 \theta$.

The dependence of the moment on the orientation of the sample is plotted in the inset to Fig. 3 and shows a $sin^2 \theta$ dependence. This establishes that only the component of the field perpendicular to the plates induces a magnetic response (leading to a factor of sin $\theta$ in the susceptibility, where $\theta$ is the angle between the field and the normal to the plates), as expected. The second factor of sin $\theta$ appears because the detector measures the component of the response in the direction of the applied field.

In order to extract the effective permeability of the metamaterial, we need to establish the mapping between this quantity and the measured magnetic moment. This mapping naturally depends on the geometry of the sample, and can be expressed in terms of a demagnetization factor. These have been tabulated[12] for rectangular prisms with isotropic permeability, but our samples are highly anisotropic and new calculations are therefore required.

When the permeability is isotropic, the divergence of the magnetization is zero everywhere within the sample, and one need only consider the surfaces in order to calculate the demagnetizing field. This is not the case when the permeability is anisotropic; we then require full volume calculations, which are computationally demanding. However, Fig. 3 shows that there is no response to a field parallel to the plates, and the parallel components of the permeability therefore take the value 1; thus we need only deal with the perpendicular component, $\mu_{perp}$, which significantly reduces the complexity of the problem. We applied the discrete dipole approximation[13] to derive the total moment and demagnetization factor as a function of $\mu_{perp}$ for our sample geometries. In this model, the system is represented as a collection of dipole moments; the strength of each moment depends (via the permeability) on the magnitude of the local field, which includes the external field and the field due to the other dipoles. The problem is thus reduced to a set of coupled linear equations. The results for the anisotropic systems lie within a few per cent of those obtained in the isotropic case.

Simulations of arrays of superconducting elements have been carried out before, in the context of strips designed for high-current applications[14]. These differ from our simulations in several respects: the dimensionality is lower (the strips are infinitely long); the actual system is simulated, rather than an effective medium model; and the aim was to calculate the susceptibility of only the superconducting elements, rather than the whole system.

We use the calculated demagnetizing factors to determine the effective permeability of our samples (Table 1).

| Length of side of plates, $l$ ($\mu$m) | Number of layers | Magnetic moment over field ($10^{-4}$ emu Oe$^{-1}$) | Demagnetizing factor | Effective permeability (predicted) | Effective permeability (measured) |
| --- | --- | --- | --- | --- | --- |
| 133 | 9 | $-1.05 \pm 0.11$ | $0.483 \pm 0.005$ | 0.64 | $0.58 \pm 0.04$ |
| 153 | 6 | $-0.97 \pm 0.10$ | $0.580 \pm 0.005$ | 0.48 | $0.49 \pm 0.05$ |
| 163 | 9 | $-2.10 \pm 0.21$ | $0.489 \pm 0.005$ | 0.23 | $0.31 \pm 0.06$ |
| 167 | 9 | $-3.70 \pm 0.37$ | $0.504 \pm 0.005$ | 0.0 | $0.04 \pm 0.06$ |

**Table 1** The effective permeability of the metamaterial samples. The in-plane lattice constant $d$ is 167 $\mu$m and the gap size is $d - l$. The final sample with $l = 167$ $\mu$m is the control, which has no gaps. The most significant contribution to the error in the measured effective permeability comes from the experimentally determined magnetic moment, which is subject to an uncertainty of up to 10%. The quoted error also allows for the uncertainty in the position of the boundary of the metamaterial (up to half a layer). Careful analysis of the finite-size errors in the discrete dipole approximation allowed us to obtain demagnetizing factors accurate to within 1%.

The table compares the measured values with those predicted by band-structure simulations[4]; there is good agreement between the two. Qualitatively similar results were obtained for the sample made from the flexible substrate, demonstrating that a large area, non vacuum process has been identified for the purposes of application of this work. In conclusion, we have constructed the first DC metamaterial, which demonstrates a tuneable, highly anisotropic diamagnetic response, albeit at low temperature and operating in low applied magnetic fields. Most of the experimental difficulty lies in fabricating samples of sufficiently small size to fit inside a magnetometer. When this size constraint is lifted, structures of higher quality can be produced with greater ease. A metamaterial with an anisotropic diamagnetic response is an essential ingredient of a cloak for static magnetic fields.